\documentclass[11pt]{article}

 \newbox
\pippobox

\begin{document}

\title{{\Large \textbf{Equilibrium points of the tilted perfect fluid Bianchi VI$_h$
state space}}}
\author{Pantelis S. Apostolopoulos$\thanks{%
E-mail: \emph{papost@phys.uoa.gr}.}$ \\
{\small \textit{University of Athens, Department of Physics,}}\\
{\small \textit{Nuclear and Particle Physics Section},}\\
{\small \textit{Panepistemiopolis, Zografos 157 71, Athens, Greece}}}
\maketitle

\begin{abstract}
We present the full set of evolution equations for the spatially homogeneous
cosmologies of type VI$_h$ filled with a tilted perfect fluid and we provide
the corresponding equilibrium points of the resulting dynamical state space.
It is found that only when the group parameter satisfies $h>-1$ a
self-similar solution exists. In particular we show that for $h>-{\frac 19}$
there exists a self-similar equilibrium point provided that $\gamma \in
\left( {\frac{2(3+\sqrt{-h})}{5+3\sqrt{-h}}},{\frac 32}\right) $ whereas for 
$h<-{\frac 19}$ the state parameter belongs to the interval $\gamma \in
\left( 1,{\frac{2(3+\sqrt{-h})}{5+3\sqrt{-h}}}\right) $. This family of new
exact self-similar solutions belongs to the subclass $n_\alpha ^\alpha =0$
having non-zero vorticity. In both cases the equilibrium points have a five
dimensional stable manifold and may act as future attractors at least for
the models satisfying $n_\alpha ^\alpha =0$. Also we give the exact form of
the self-similar metrics in terms of the state and group parameters. As an
illustrative example we provide the explicit form of the corresponding
self-similar radiation model ($\gamma={\frac 43}$), parametrised by the
group parameter $h$. Finally we show that there are no tilted self-similar
models of type III and irrotational models of type VI$_h$.\newline
\end{abstract}

\section{Introduction}

\setcounter{equation}{0}

Although on a sufficiently large observational scale, the present state of
the Universe is described by the Friedmann-Lema\^{i}tre (FL) model which is
isotropic and spatially homogeneous, there are potential problems mainly
regarding the observed \emph{local} structures of our ``lumpy'' Universe
which cannot be explained within the class of FL models. Therefore more
general cosmological models, which in some dynamical sense, are ``close'' to
FL but not isotropic in local scale, can be used in order to answer many
open and important questions. For example it is of interest to understand
the presence, the form and the evolution of small (local) density and
expansion anisotropies in the Universe or to investigate the constraints
that measurements of temperature anisotropies are able to impose on the
curvature of space-time. In addition it is important to classify all
possible asymptotic states near the cosmological initial singularity (i.e.
near the Planck time) and into the future that are permitted by the
Einstein's Field Equations (EFE) with a view to explaining how the real
Universe may have evolved.

In this point of view the simplest (anisotropic) generalisation of the FL
universes are the Spatially Homogeneous (SH) cosmologies admitting a $G_3$
group of isometries acting on 3-dimensional spacelike hypersurfaces $%
\mathcal{C}$. From cosmological point of view, it is of importance to study
the evolution of vacuum or models filled with a gamma-law perfect fluid
matter source having an energy-momentum tensor of the form: 
\begin{equation}
T_{ab}=(\tilde{\mu}+\tilde{p})u_au_b+\tilde{p}g_{ab}  \label{Intro1}
\end{equation}
where $\tilde{\mu},\tilde{p}=\left( \gamma -1\right) \tilde{\mu}$ are the
energy density and the pressure measured by the observers comoving with
fluid velocity $u^a$ ($u^au_a=-1$). Since in SH models there is a preferred
unit timelike congruence $n^a$ ($n^an_a=-1$) normal to the spatial
foliations $\mathcal{C}$ we can divide them into \emph{non-tilted} \cite
{Ellis-MacCallum} and \emph{tilted }\cite{King-Ellis} models according to
whether the fluid velocity $u^a$ is parallel or not to the timelike
direction $n^a$.

In the last two decades, the study of SH models is heavily based on the
qualitative analysis of the resulting system of the (induced) first order
ordinary differential equations. Using the so called \emph{orthonormal frame
formalism} (pioneered by Ellis \cite{Ellis-MacCallum}) which is based on
choosing a frame tetrad invariant under the group of isometries (thus
ensuring the spatial independency of the kinematical and dynamical
quantities of the models) and a set of expansion-normalized variables, the
evolution and constraint equations, followed from the EFE, become an
autonomous system of decoupled first order differential equations which can
be studied with the aid of the well-established theory of dynamical systems 
\cite{Wainwright-Ellis}. The evolution of a specific model is studied in the
so called \emph{dynamical state space} which represents the set of all the
physical states (at some instant of time) of the corresponding model \cite
{Wainwright-Ellis}. Under this perspective of studying SH models, \emph{%
equilibrium points} (i.e. fixed points) and their stability, of the
resulting dynamical system, play an important role in the description of the
asymptotic behavior (into the past/future as well as in the intermediate
times of their evolution) since they may represent past or future attractors
for more general models.

Vacuum and non-tilted perfect fluid SH models have been extensively studied
in the literature \cite
{Wainwright-Ellis,Wainwright-Hancock-Uggla,Wainwright1,Nilsson-Hancock-Wainwright1,Ringstrom,Horwood-Hancock-The-Wainwright,Hewitt-Horwood-Wainwright}
revealing new and important features of the SH models like e.g. the \emph{%
asymptotically self-similarity breaking} and the \emph{divergence of the
Weyl curvature} at late times for type VII$_0$ models, providing a solid
counterexample to the isotropisation conjecture (i.e. shear isotropisation
implies a corresponding result for the Weyl curvature scalar). On the other
hand it is natural to expect that the behavior of SH models will be modified
accordingly by the presence of a tilted fluid velocity leading also to new
interesting phenomena.

The first step of qualitative analyzing tilted models has been done for
Bianchi type II models \cite{Hewitt-Bridson-Wainwright}. In particular it
has been shown that $\gamma -$law tilted perfect fluid models, are future
asymptotic to the Collins-Stewart non-tilted model \cite{Collins-Stewart1}
when $\frac 23<\gamma \leq \frac{10}7$, consequently these models do not
isotropise and the angle of tilt becomes negligible at late times. At the
value $\gamma =\frac{10}7$ the tilt destabilise the Collins-Stewart model
and there is an exchange of stability with the self-similar equilibrium
point in which $\gamma \in \left( \frac{10}7,\frac{14}9\right) $.
Furthermore at the value $\gamma =\frac{14}9$ there is a second bifurcation
between the equilibria $\left( \frac{10}7,\frac{14}9\right) $ and $\left( 
\frac{14}9,2\right) $ and exhibits the property of the \emph{asymptotically
extreme tilt} for models where the state parameter $\gamma $ belongs to the
interval $\left( \frac{14}9,2\right) $.

Recently it was shown that the self-similar equilibrium points of Bianchi
type VI$_0$ models play a similar role in the asymptotic behaviour of
generic models. For example it was found \cite{Apostol12} that at the value $%
\gamma =\frac 65$ the tilt destabilise the Collins solution \cite{Collins2}
and a family of models satisfying $n_\alpha ^\alpha =0$ are future
asymptotic to the Rosquist and Jantzen self-similar model \cite
{Apostol12,Rosquist-Jantzen1} for $\gamma \in \left( \frac 65,\frac
32\right) $. However it has been shown that generic models (i.e. those
satisfying $n_\alpha ^\alpha \neq 0$) are not asymptotically self-similar 
\cite{Apostol12} and may be extreme tilted at late times for $\frac
65<\gamma<2$ \cite{Hervik1}.

In the case of class B tilted models less information is available due to
the increased complexity of the evolution equations. Recently the whole
family of Bianchi class B models has been studied and some results
concerning the stability of the non-tilted equilibrium points have been
given \cite{Barrow-Hervik} (see also \cite {Hervik-Hoogen-Coley, Coley-Hervik}).

Motivated from the above facts, the goal of this work is to present the full
set of evolution equations, the vacuum, non-tilted and tilted equilibrium
points for the important class of Bianchi type VI$_h$ models and interpret,
for some of them, their geometric and dynamical properties.

An outline of the paper is as follows: section 2 reviews and presents the
basic results concerning the set of equations which describes the dynamics
of the tilted perfect fluid SH models. By specialising to the case of
Bianchi type VI$_h$ models, we provide the complete set of the evolution
equations and we identify the resulting dynamical state space. In section 3
we find the complete set of equilibrium points and for the case which are
represented by self-similar models, we give all tilted perfect fluid models
admitting a proper Homothetic Vector Field (HVF). Finally in section 4 we
summarise and discuss the implications some of the obtained results.

Throughout the following conventions have been used: spatial frame indices
are denoted by lower Greek letters $\alpha ,\beta ,...=1,2,3$, lower Latin
letters denote space-time indices $a,b,...=0,1,2,3$ and we use geometrised
units such that $8\pi G=c=1$.

\section{Dynamical state space of tilted perfect fluid Bianchi type VI$_h$
models}

\setcounter{equation}{0}

In SH tilted perfect fluid models, the autonomous differential equation
governing their evolution can be written in the form: 
\begin{equation}
\frac{d\mathbf{x}}{d\tau }=f\left( \mathbf{x}\right)  \label{diffequat1}
\end{equation}
where $\mathbf{x}$ is the \emph{state vector} representing the set of all
the physical variables that describe the dynamics of the corresponding
model, $f(\mathbf{x})$ is a polynomial function of the state vector and $%
\tau $ is the dimensionless time variable defined by: 
\begin{equation}
\frac{dt}{d\tau }=\frac 1H,\quad \frac{dH}{d\tau }=-\left( 1+q\right) H
\label{diffequat2}
\end{equation}
where $q,H$ are the deceleration and Hubble parameter respectively.

In \cite{Hewitt-Bridson-Wainwright} and using the orthonormal frame
approach, the EFE are reformulated in terms of the components of the shear
tensor of the normal timelike congruence $n^a$, the spatial curvature of the
orbits of the $G_3$ isometry group and the spatial part of the tilted fluid
velocity $u^a$. The evolution equations for the type VI$_h$ models can be
found by specialising the set of equations given in \cite
{Hewitt-Bridson-Wainwright}:

\begin{equation}
\Sigma _{\alpha \beta }^{\prime }=-\left( 2-q\right) \Sigma _{\alpha \beta
}+2\epsilon _{\hspace{0.3cm}(\alpha }^{\mu \nu }\Sigma _{\beta )\mu }R_\nu
-S_{\alpha \beta }+\Pi _{\alpha \beta }  \label{evol1}
\end{equation}
\begin{equation}
N_{\alpha \beta }^{\prime }=qN_{\alpha \beta }+2\Sigma _{(\alpha }^{\hspace{%
0.2cm}\mu }N_{\beta )\mu }+2\epsilon _{\hspace{0.3cm}(\alpha }^{\mu \nu
}N_{\beta )\mu }R_\nu  \label{evol2}
\end{equation}
\begin{equation}
A_\alpha ^{\prime }=qA_\alpha -\Sigma _\alpha ^{\hspace{0.2cm}\mu }A_\mu
+\epsilon _{\hspace{0.3cm}\alpha }^{\mu \nu }A_\mu R_\nu  \label{evol3}
\end{equation}
\begin{equation}
\Omega ^{\prime }=\Omega G^{-1}\left[ 2Gq-\left( 3\gamma -2\right) -\left(
2-\gamma \right) v^2-\gamma \Sigma _{\mu \nu }v^\mu v^\nu +2\gamma A_\mu
v^\mu \right]  \label{Bianchi1}
\end{equation}
\begin{eqnarray}
v_\alpha ^{\prime } &=&\frac{v_\alpha }{\left[ 1-\left( \gamma -1\right)
v^2\right] }\{\left( 3\gamma -4\right) \left( 1-v^2\right) +\left( 2-\gamma
\right) \Sigma _{\gamma \delta }v^\gamma v^\delta +  \nonumber \\
&&  \nonumber \\
&&+\left[ \left( 2-\gamma \right) -\left( \gamma -1\right) \left(
1-v^2\right) \right] A_\beta v^\beta \}-\Sigma _\alpha ^{\hspace{0.1cm}\beta
}v_\beta +  \nonumber \\
&&  \nonumber \\
&&+\epsilon _\alpha ^{\hspace{0.2cm}\mu \nu }\left( -R_\mu +N_\mu ^{\hspace{%
0.1cm}\delta }v_\delta \right) v_\nu -v^2A_\alpha  \label{Bianchi2}
\end{eqnarray}
where a prime denotes derivative w.r.t. $\tau$.

The above system is subjected to the algebraic constraints: 
\begin{equation}
\Omega =1-\Sigma ^2-K  \label{algebraic1}
\end{equation}
\begin{equation}
3\gamma G^{-1}\Omega v_\alpha =3\Sigma _\alpha ^{\hspace{0.2cm}\beta
}A_\beta -\epsilon _\alpha ^{\hspace{0.2cm}\mu \nu }\Sigma _\mu ^{\hspace{%
0.2cm}\beta }N_{\beta \nu }  \label{algebraic2}
\end{equation}
where we have set 
\begin{equation}
G=1+\left( \gamma -1\right) v^2  \label{algebraic3}
\end{equation}
and the deceleration parameter is given by the relation: 
\begin{eqnarray}
q &=&2\Sigma ^2+\frac 12G^{-1}\Omega \left[ \left( 3\gamma -2\right) \left(
1-v^2\right) +2\gamma v^2\right] =  \nonumber \\
&=&2\left( 1-K\right) -\frac 12G^{-1}\Omega \left[ 3\left( 2-\gamma \right)
\left( 1-v^2\right) +2\gamma v^2\right] .  \label{algebraic4}
\end{eqnarray}
Using the freedom of a time-dependent spatial rotation, we may choose the
orthonormal tetrad to be the eigenframe of $N_{\alpha \beta }$ therefore the
contracted form of Jacobi identities $N_{\alpha \beta }A^\beta =0$ implies:

\begin{equation}
N_{\alpha \beta }=\left( 
\begin{array}{lll}
0 & 0 & 0 \\ 
0 & N_2 & 0 \\ 
0 & 0 & N_3
\end{array}
\right) ,\qquad A_\alpha =A_1\delta _\alpha ^1.  \label{def1}
\end{equation}
The evolution equation of $N_{\alpha \beta }$ can be used to express the
angular velocity $R_\alpha $ of the spatial frame in terms of the shear
variables: 
\begin{equation}
R_1=\frac{N_2+N_3}{N_2-N_3}\Sigma _{23},\quad R_2=\Sigma _{13},\quad
R_3=-\Sigma _{12}.  \label{fermi1}
\end{equation}
In addition equations (\ref{evol2}) and (\ref{evol3}) have a first integral
which is used to express the component $A_1$ in the well known form: 
\begin{equation}
A_1^2=hN_2N_3.  \label{first_integral}
\end{equation}
Following \cite{Hewitt-Bridson-Wainwright} we introduce the shear variables: 
\begin{equation}
\Sigma _{+}=\frac 12\left( \Sigma _{22}+\Sigma _{33}\right) ,\quad \Sigma
_{-}=\frac 1{2\sqrt{3}}\left( \Sigma _{22}-\Sigma _{33}\right)
\label{shear1}
\end{equation}
\begin{equation}
\Sigma _1=\frac 1{\sqrt{3}}\Sigma _{23},\quad \Sigma _3=\frac 1{\sqrt{3}%
}\Sigma _{12},\quad \Sigma _{13}\rightarrow \frac 1{\sqrt{3}}\Sigma _{13}.
\label{shear2}
\end{equation}
In the case of type VI$_h$ models we have $h<0$ and $N_2N_3<0$. With these
identifications we obtain the following set of evolution equations for the
basic expansion-normalised variables $\mathbf{x}=\left( \Sigma _{+},\Sigma
_{-},\Sigma _1,\Sigma _3,\Sigma _{13},N_2,N_3,v_\alpha \right) $: 
\begin{equation}
\Sigma _{+}^{\prime }=-\left( 2-q\right) \Sigma _{+}-\frac{\left(
N_2-N_3\right) ^2-18\left( \Sigma _{13}^2+\Sigma _3^2\right) }6-\frac{\Omega
\gamma \left( 2v_1^2-v_2^2-v_3^2\right) }{2G}  \label{statevector1}
\end{equation}
\begin{eqnarray}
\Sigma _{-}^{\prime } &=&-\left( 2-q\right) \Sigma _{-}-\frac{\sqrt{3}\left(
N_2^2-N_3^2\right) }6+\frac{4\sqrt{3}N_3\Sigma _1^2}{N_2-N_3}+  \nonumber \\
&&  \nonumber \\
&&+\sqrt{3}\left( 2\Sigma _1^2-\Sigma _{13}^2+\Sigma _3^2\right) +\frac{%
\sqrt{3}\Omega \gamma \left( v_2^2-v_3^2\right) }{2G}  \label{statevector2}
\end{eqnarray}
\begin{eqnarray}
\Sigma _1^{\prime } &=&-\left[ \frac{4\sqrt{3}N_3\Sigma _{-}}{N_2-N_3}%
-q+2\left( \sqrt{3}\Sigma _{-}+1\right) \right] \Sigma _1+  \nonumber \\
&&  \nonumber \\
&&+\sqrt{3}\frac{\sqrt{hN_2N_3}\left( N_3-N_2\right) +6\Sigma _{13}\Sigma _3}%
3+\frac{\sqrt{3}v_2v_3\Omega \gamma }G  \label{statevector3}
\end{eqnarray}
\begin{equation}
\Sigma _3^{\prime }=\left( q-\sqrt{3}\Sigma _{-}-3\Sigma _{+}-2\right)
\Sigma _3+\frac{2\sqrt{3}N_3\Sigma _{13}\Sigma _1}{N_2-N_3}+\frac{\sqrt{3}%
v_1v_2\Omega \gamma }G  \label{statevector4}
\end{equation}
\begin{equation}
\Sigma _{13}^{\prime }=\left( q+\sqrt{3}\Sigma _{-}-3\Sigma _{+}-2\right)
\Sigma _{13}-\frac{2\sqrt{3}N_2\Sigma _3\Sigma _1}{N_2-N_3}+\frac{\sqrt{3}%
v_1v_3\Omega \gamma }G  \label{statevector5}
\end{equation}
\begin{equation}
N_2{}^{\prime }=\left( q+2\sqrt{3}\Sigma _{-}+2\Sigma _{+}\right) N_2
\label{statevector6}
\end{equation}
\begin{equation}
N_3{}^{\prime }=\left( q-2\sqrt{3}\Sigma _{-}+2\Sigma _{+}\right) N_3
\label{statevector7}
\end{equation}
and the evolution equation (\ref{Bianchi2}) for the frame components of the
tilted fluid velocity.

The algebraic constraint (\ref{algebraic2}) reads: 
\begin{equation}
\sqrt{3}\Sigma _1\left( N_3-N_2\right) +6\sqrt{hN_2N_3}\Sigma _{+}+\frac{%
3v_1\Omega \gamma }G=0  \label{constraints1}
\end{equation}
\begin{equation}
\sqrt{3}\Sigma _{13}N_3+3\sqrt{3}\sqrt{hN_2N_3}\Sigma _3-\frac{3v_2\Omega
\gamma }G=0  \label{constraints2}
\end{equation}
\begin{equation}
\sqrt{3}\Sigma _3N_2-3\sqrt{3}\sqrt{hN_2N_3}\Sigma _{13}+\frac{3v_3\Omega
\gamma }G=0.  \label{constraints3}
\end{equation}
We note that the shear scalar $\Sigma ^2=\frac{\Sigma ^{\alpha \beta }\Sigma
_{\alpha \beta }}6$ and the spatial curvature $K$ are:\ 
\begin{equation}
\Sigma ^2=\Sigma _{+}^2+\Sigma _{-}^2+\Sigma _1^2+\Sigma _3^2+\Sigma _{13}^2
\label{defshearscalar}
\end{equation}
\begin{equation}
K=hN_2N_3+\frac{\left( N_2-N_3\right) ^2}{12}  \label{defcurvature}
\end{equation}
therefore the inequality $\Omega \geq 0$ and the constraint (\ref{algebraic1}%
) imply that the state space $\mathcal{D}\subset \mathbf{R}^{7}$ is bounded
(we recall that $N_2N_3<0$).

\section{Determination of the equilibrium points}

\setcounter{equation}{0}

Equilibrium points of the autonomous differential equation (\ref{diffequat1}%
) play an important role in the evolution of the SH models since they
determine various stable and unstable invariant submanifolds of the state
space $\mathcal{D}$. These points can be found from the solution of the
algebraic equations $f(\mathbf{x})=0$ and (\ref{constraints1})-(\ref
{constraints3}) which we now list in the following subsections. We note that
the case of type III models is included by setting (whenever is appropriate) 
$h=-1$.

\subsection{Vacuum Equilibrium Points}

\textit{1. Kasner Circle }$\mathcal{K}$\textit{\ }\cite
{Hewitt-Bridson-Wainwright}
\[
N_2=N_3=0,\qquad v^\alpha v_\alpha =0,\qquad \Sigma ^2=1 
\]
\[
\Sigma _{+}^2+\Sigma _{-}^2=1,\qquad \Sigma _1=\Sigma _{13}=\Sigma
_3=0,\qquad q=2. 
\]
\\
\textit{2. Kasner Line with tilt }$\mathcal{K}_{tilt}^{\pm }$%
\textit{\ }\cite{Hewitt-Bridson-Wainwright}
\[
N_2=N_3=0,\quad v^\alpha v_\alpha =v_3^2<1,\quad v_1=v_2=0,\quad \Sigma ^2=1 
\]
\[
\Sigma _{+}=\sqrt{3}\Sigma _{-}+3\gamma -4,\qquad \Sigma _{-}=-\frac{\sqrt{3}%
\left( 3\gamma -4\right) \pm \sqrt{3\left( 2-\gamma \right) \left( 3\gamma
-2\right) }}4 
\]
\[
\Sigma _1=\Sigma _3=\Sigma _{13}=0,\qquad q=2,\qquad \frac 23\leq \gamma
\leq 2. 
\]
\\
\textit{3. Kasner Circle with extreme tilt }$\mathcal{K}%
_{extreme}$\textit{\ }\cite{Hewitt-Bridson-Wainwright}
\[
N_2=N_3=0,\quad v^\alpha v_\alpha =v_3^2=1,\quad v_1=v_2=0,\quad \Sigma ^2=1 
\]
\[
\Sigma _{+}^2+\Sigma _{-}^2=1,\qquad \Sigma _1=\Sigma _3=\Sigma
_{13}=0,\qquad q=2,\qquad 0<\gamma <2. 
\]
\\
\textit{4. Collins Vacuum Plane Wave Arc }$\mathcal{L}(VI_h)$ ($%
h\neq -\frac 19$) \cite{Wainwright-Ellis}
\begin{eqnarray*}
-12hN_2N_3 &=&2\sqrt{3}\sqrt{-\left( N_2-N_3\right) ^2+3}+\left(
N_2-N_3\right) ^2-6,\qquad v^\alpha v_\alpha =0, \\
&&
\end{eqnarray*}
\[
\Sigma ^2=\frac{\left[ \sqrt{3}+\sqrt{-\left( N_2-N_3\right) ^2+3}\right]
^2\left[ 12hN_2N_3+\left( N_2-N_3\right) ^2\right] }{12\left( N_2-N_3\right)
^2} 
\]
\[
\Sigma _{+}=-\frac{\sqrt{3}\left[ \sqrt{3}+\sqrt{-\left( N_2-N_3\right) ^2+3}%
\right] }6,\qquad \Sigma _1=\frac{2\sqrt{3hN_2N_3}\Sigma _{+}}{\left(
N_2-N_3\right) },\qquad 
\]
\[
\Sigma _{-}=\Sigma _{13}=\Sigma _3=0,\qquad q=\frac{\sqrt{3}\left( \sqrt{%
-\left( N_2-N_3\right) ^2+3}+\sqrt{3}\right) }3 
\]
\vspace{0.3cm} 
\[
0<\gamma <2. 
\]
\\
\textit{5. Vacuum plane wave with tilt\footnote{%
It appears that this form of the Collins type VI$_h$ plane wave solution has
been also given in \cite{Coley-Hervik} using, however, a different notation. 
} }$\mathcal{M}_{tilt}^{\pm }(VI_h)$ ($h\neq -\frac 19$)
\[
N_2=N_2,\quad N_3=N_3,\quad v^\alpha v_\alpha =\frac{\left[ \sqrt{-\left(
N_2-N_3\right) ^2+3}\mp \sqrt{3}\left( 3\gamma -5\right) \right] ^2}{%
12hN_2N_3\left( \gamma -1\right) ^2}, 
\]
\vspace{0.5cm} 
\[
\Sigma ^2=\frac{\left[ \sqrt{-\left( N_2-N_3\right) ^2+3}\pm \sqrt{3}\right]
^2\left[ 12hN_2N_3+\left( N_2-N_3\right) ^2\right] }{12\left( N_2-N_3\right)
^2} 
\]
\vspace{0.5cm} 
\[
\Sigma _{+}=\frac{\sqrt{3}\left[ \mp \sqrt{-\left( N_2-N_3\right) ^2+3}-%
\sqrt{3}\right] }6,\qquad \Sigma _{-}=\Sigma _{13}=\Sigma
_3=\upsilon_2=\upsilon_3=0 
\]
\vspace{0.5cm} 
\[
\Sigma _1=\frac{6\Sigma _{+}\sqrt{hN_2N_3}}{\left( N_2-N_3\right) \sqrt{3}} 
\]
\vspace{0.5cm} 
\[
v_1=\frac{\sqrt{3}\left[ \pm \sqrt{-\left( N_2-N_3\right) ^2+3}-\sqrt{3}%
\left( 3\gamma -5\right) \right] }{6\sqrt{hN_2N_3}\left( 1-\gamma \right) }, 
\]
\vspace{0.5cm} 
\[
q=\frac{\sqrt{3}\left[ \sqrt{3}\pm \sqrt{-\left( N_2-N_3\right) ^2+3}\right] 
}3 
\]
\vspace{0.5cm} 
\[
h=\frac{\left( N_2-N_3\right) ^4}{12N_2N_3\left[ \pm 2\sqrt{3}\sqrt{-\left(
N_2-N_3\right) ^2+3}-\left( N_2-N_3\right) ^2+6\right] }. 
\]
\vspace{0.5cm}

\noindent We remark that the state parameter $\gamma$ is constrained via the
inequality $1-v^2>0$.\newline
\newline
\textit{6. Vacuum plane wave with extreme tilt }$\mathcal{M}_{extreme}(VI_h)$
($h\neq -\frac 19$).\newline
Same as the case $\mathcal{M}_{tilt}^{\pm }(VI_h)$. However the state
parameter can take any value in the interval $\left( 0,2\right) $.

\subsection{Non Vacuum Equilibrium Points}

\textit{1. Flat Friedmann-Lema\^{i}tre Equilibrium Point }$\mathcal{F}$%
\textit{\ }\cite{Hewitt-Bridson-Wainwright}
\[
N_2=N_3=0,\qquad v^\alpha v_\alpha =0,\qquad \Sigma ^2=0 
\]
\[
\Sigma _{+}=\Sigma _{-}=\Sigma _1=\Sigma _{13}=\Sigma _3=0,\qquad q=\frac{%
3\gamma -2}2,\qquad \Omega =1 
\]
\[
0<\gamma<2. 
\]
\\
\textit{2. Collins-Stewart type II non-tilted Equilibrium
Point }$\mathcal{CS}(II)$\textit{\ }\cite{Wainwright-Ellis}
\[
N_2=0,N_3=\frac{3\sqrt{\left( 2-\gamma \right) \left( 3\gamma -2\right) }}%
4,\qquad v^\alpha v_\alpha =0,\qquad \Sigma ^2=\frac{\left( 3\gamma
-2\right) ^2}{64} 
\]
\[
\Sigma _{+}=\frac{2-3\gamma }{16},\qquad \Sigma _{-}=\frac{\sqrt{3}\left(
3\gamma -2\right) }{16},\qquad \Sigma _1=\Sigma _{13}=\Sigma _3=0 
\]
\[
q=\frac{3\gamma -2}2,\qquad \frac 23<\gamma <2,\qquad \Omega =\frac{3\left(
6-\gamma \right) }{16}. 
\]
\\
\textit{3. Hewitt type II tilted Equilibrium Point }$\mathcal{%
P}_{tilt}(II)$\textit{\ }\cite{Hewitt-Bridson-Wainwright}
\[
N_2=0,N_3=3\sqrt{\frac{\left( \gamma -2\right) \left( 3\gamma -4\right)
\left( 5\gamma -4\right) }{18-17\gamma }},\quad v^\alpha v_\alpha =\frac{%
\left( 3\gamma -4\right) \left( 7\gamma -10\right) }{\left( 11\gamma
-10\right) \left( 5\gamma -4\right) }, 
\]
\[
\Sigma ^2=\frac{\left( 3\gamma -4\right) \left( 9\gamma ^2-20\gamma
+12\right) }{17\gamma -18},\quad \Sigma _{+}=\frac{9\gamma -14}8,\quad
\Sigma _{-}=\frac{\sqrt{3}\left( 5\gamma -6\right) }8, 
\]
\[
\Sigma _{13}=\sqrt{\frac{3\left( \gamma -2\right) \left( 7\gamma -10\right)
\left( 11\gamma -10\right) }{16\left( 18-17\gamma \right) }},\quad \Sigma
_1=\Sigma _3=0, 
\]
\[
\Omega =\frac{3\left( 2-\gamma \right) \left( 21\gamma ^2-24\gamma +4\right) 
}{4\left( 17\gamma -18\right) } 
\]
\[
v_2=\sqrt{\frac{\left( 3\gamma -4\right) \left( 7\gamma -10\right) }{\left(
11\gamma -10\right) \left( 5\gamma -4\right) }},\qquad q=\frac{3\gamma -2}%
2,\qquad \frac{10}7<\gamma <2. 
\]
\\
\textit{4. Type II Line of tilted Equilibrium Points }$\mathcal{L}_{tilt}(II) $\textit{\ }\cite{Hewitt-Bridson-Wainwright}
\[
N_2=\sqrt{\frac{2\left( 27b^2+2\right) \left( 17-54b^2\right) }{57}}%
,N_3=0,\quad v^\alpha v_\alpha =\frac{6\left( 27b^2+1\right) \left(
27b^2+2\right) }{\left( 54b^2-17\right) \left( 81b^2-32\right) },
\]
\[
\Sigma ^2=\frac{\left( 2-3b^2\right) \left( 27b^2+2\right) }{19},\quad
\Sigma _3=-\frac{\sqrt{\left( 27b^2+1\right) \left( 32-81b^2\right) }}{3%
\sqrt{57}},\quad \Sigma _{13}=b,
\]
\[
\Sigma _{-}=-\frac{2\sqrt{3}}9,\quad \Sigma _{+}=\Sigma _1=0,\quad v_3=\sqrt{%
\frac{6\left( 27b^2+1\right) \left( 27b^2+2\right) }{\left( 17-54b^2\right)
\left( 32-81b^2\right) }}
\]
\[
q=\frac 43,\qquad \Omega =\frac{2916b^4-1215b^2+236}{342},\qquad \left|
b\right| <\frac 2{3\sqrt{3}},\quad \gamma =\frac{14}9.
\]
\\
\textit{5. Type II Extreme tilted Equilibrium Point }$\mathcal{P%
}_{extreme}(II)$\textit{\ }\cite{Hewitt-Bridson-Wainwright}
\[
N_2=\frac{6\sqrt{19}}{19},N_3=0,\quad v^\alpha v_\alpha =1, 
\]
\[
\Sigma ^2=\frac{28}{57},\quad \Sigma _3=-\frac{10\sqrt{57}}{171},\quad
\Sigma _{13}=\frac{2\sqrt{3}}3, 
\]
\[
\Sigma _{-}=-\frac{2\sqrt{3}}9,\quad \Sigma _{+}=\Sigma _1=0,\quad v_3=1 
\]
\[
q=\frac 43,\qquad \Omega =\frac{20}{57},\qquad 0<\gamma <2. 
\]
\\
\textit{6. Collins Type VI}$_h$\textit{\ non-tilted
Equilibrium Point }$\mathcal{C}^{\pm }(VI_h)$\textit{\ }\cite
{Wainwright-Ellis}
\[
N_2=\frac{3\sqrt{\left( 2-\gamma \right) \left( 3\gamma -2\right) }}%
4,N_3=-N_2,\quad v^\alpha v_\alpha =0,
\]
\[
\Sigma ^2=\frac{\left( 3\gamma -2\right) ^2\left( 1-3h\right) }{16},\quad
\Sigma _3=\Sigma _{13}=\Sigma _{-}=0,
\]
\[
\Sigma _{+}=\frac{2-3\gamma }4,\quad \Sigma _1=\frac{\pm \sqrt{3}\left(
2-3\gamma \right) \sqrt{-h}}4,
\]
\[
q=\frac{3\gamma -2}2,\qquad \Omega =\frac{3\left[ h\left( 3\gamma -2\right)
-\gamma +2\right] }4,\qquad \frac 23\leq \gamma \leq \frac{2\left(
1-h\right) }{\left( 1-3h\right) }.
\]
\\
\textit{7. Type VI}$_h$\textit{\ tilted Equilibrium Point }$%
\mathcal{C}_{tilt}(VI_h)$ ($h=-h_1^2$ and $h_1\neq -1/3$): 
\begin{eqnarray*}
&&-2h_1\gamma \left[ -\left( 3h_1+1\right) \right] ^{1/2}\left\{
3h_1^2\left( 4q\gamma -\beta \right) -4h_1\left[ q\left( 5\gamma -4\right)
+2\left( \gamma -2\right) \right] +3\beta \right\} ^{1/2}N_3 \\
&& \\
&=&\sqrt{3}\{6h_1^4\gamma \left[ \left( q+1\right) \left( 2q\gamma -\beta
\right) \left( 6q\left( \gamma -1\right) +5\gamma -6\right) \right]
-h_1^3[\gamma ^3\left( 120q^3+16q^2-28q+49\right) + \\
&& \\
&&+\left( q+1\right) \left( -2\gamma ^2(144q^2+32q+23)+60\gamma
(q+1)(4q+1)-72(q+1)^2\right) ]+ \\
&& \\
&&+h_1^2\beta [\gamma ^2(8q^2-66q-41)+(q+1)\left( 4\gamma
(q+18)-12(q+1)\right) ]+ \\
&& \\
&&h_1\beta ^2\left[ 2q\left( 4\gamma -5\right) +13\gamma -10\right] -\beta
^3\}^{1/2}
\end{eqnarray*}
\vspace{0.5cm} 
\[
v^\alpha v_\alpha =\frac{3\beta ^2\left\{ h_1^2\left[ 6q\left( \gamma
-1\right) +5\gamma -6\right] -2h_1\left[ q\left( 3\gamma -2\right) +\gamma
-2\right] +\beta \right\} }{4h_1^2N_3^2\gamma ^2\left\{ 3h_1^2\left(
4q\gamma -\beta \right) -4h_1\left[ q\left( 5\gamma -4\right) +2\left(
\gamma -2\right) \right] +3\beta \right\} }
\]
\vspace{0.5cm} 
\[
\Sigma _3=-\frac{\sqrt{3}\left\{ 2h_1N_3\left[ \zeta +v_1^2\gamma ^2\left(
\gamma -1\right) \right] +2N_3\zeta +v_1\gamma ^2\left( \beta +2-q\right)
\right\} }{12\zeta ^{1/2}\gamma },
\]
\vspace{0.5cm} 
\[
\Sigma _{13}=\Sigma _3,\quad \Sigma _1=\frac{\sqrt{3}\left[ h_1\left(
q+1\right) \left( 5\gamma -6\right) -\beta \right] }{6h_1\gamma },\quad
\Sigma _{+}=-\frac q2,\quad \Sigma _{-}=0
\]
\vspace{0.5cm} 
\[
\Omega =-\frac{\left[ 3h_1^2q\gamma +h_1\left( q+1\right) \left( 6-5\gamma
\right) +\beta \right] \left\{ 4h_1^2N_3^2\left[ 2\zeta \left( \gamma
-1\right) +\gamma ^2\right] +\left( \gamma -1\right) \beta ^2\right\} }{%
6h_1^2\gamma ^3\beta }
\]
\vspace{0.5cm} 
\[
v_1=-\frac \beta {2h_1N_3\gamma },\quad v_2=v_3=\zeta ^{1/2}\gamma ^{-1}
\]
\vspace{0.5cm} 
\[
q=\frac{A+B}\Lambda 
\]
where we have set:\ 
\begin{eqnarray*}
A &=&3h_1\gamma \left| 1+3h_1\right| \left| h_1\left( 3\gamma -2\right)
-5\gamma +6\right| \times  \\
&& \\
&&\times \{\left( \gamma -1\right) [h_1^2\left( \gamma -1\right) \left(
7\gamma -6\right) ^2+2h_1\left( \gamma -2\right) \left( 27\gamma ^2-37\gamma
+6\right) + \\
&& \\
&&+\left( \gamma -2\right) ^2\left( 9\gamma -1\right) ]\}^{1/2}
\end{eqnarray*}
\vspace{0.5cm} 
\begin{eqnarray*}
B &=&18h_1^4\gamma ^2\left( \gamma -1\right) \left( 3\gamma -2\right)
+3h_1^3\left( 66\gamma ^4-427\gamma ^3+808\gamma ^2-588\gamma +144\right) -
\\
&& \\
&&-3h_1^2\left( 106\gamma ^4-595\gamma ^3+1200\gamma ^2-1052\gamma
+336\right) - \\
&& \\
&&-3h_1\left( 90\gamma ^4-499\gamma ^3+984\gamma ^2-844\gamma +272\right)
+\left( 3\gamma -2\right) \left( 35\gamma -36\right) \left( \gamma -2\right)
]
\end{eqnarray*}
\vspace{0.5cm} 
\begin{eqnarray*}
\Lambda  &=&2[27h_1^4\gamma \left( \gamma -1\right) ^2\left( 3\gamma
-2\right) -18h_1^3\left( 15\gamma ^4-62\gamma ^3+93\gamma ^2-58\gamma
+12\right) + \\
&& \\
&&+3h_1^2\left( 75\gamma ^4-384\gamma ^3+704\gamma ^2-560\gamma +168\right) -
\\
&& \\
&&-6h_1\left( 30\gamma ^3-127\gamma ^2+166\gamma -68\right) +\left( 35\gamma
-36\right) \left( \gamma -2\right) ]
\end{eqnarray*}
\vspace{0.5cm} 
\[
\beta =2q-3\gamma +2,
\]
\vspace{0.5cm} 
\[
\zeta =\frac{\beta ^2\left( \gamma -2\right) \left( 1+3h_1\right) \left(
q+1\right) }{4h_1N_3^2\left\{ 3h_1^2\left( 4q\gamma -\beta \right)
-4h_1\left[ q\left( 5\gamma -4\right) +2\left( \gamma -2\right) \right]
+3\beta \right\} }
\]
\\
We note that the above solution is defined only when $h_1>-1$
and belongs to the subclass $n_\alpha ^\alpha =0$. In addition the state
parameter satisfies:\ 
\[
-1<h_1<-\frac 13\Rightarrow 1<\gamma <\frac{2\left( 3-h_1\right) }{5-3h_1}
\]
\vspace{0.5cm} 
\[
-\frac 13<h_1<0\Rightarrow \frac{2\left( 3-h_1\right) }{5-3h_1}<\gamma
<\frac 32.
\]
It was not proved possible to study the stability properties of the solution
in the full state space. However the simpler problem of the subclass of
models satisfying $n_\alpha ^\alpha =0$ can be treated. In particular a
preliminary analysis indicates that this equilibrium point has a five
dimensional stable manifold and may act as the future attractor at least for
the models satisfying $n_\alpha ^\alpha =0$.\newline
\newline
\textit{8. Type VI}$_h$\textit{\ Extreme tilted Set of Equilibrium Points }$%
\mathcal{C}_{extreme}(VI_h)$ ($h=-h_1^2$ and $h_1\neq -1/3$): 
\[
N_2=N_2,N_3=N_3,\quad v^\alpha v_\alpha =1,
\]
\vspace{0.5cm} 
\[
\Sigma ^2=\frac{24h_1\sqrt{-N_2N_3}-12h_1^2N_2N_3+\left( N_2-N_3\right) ^2+12%
}{12},
\]
\vspace{0.5cm} 
\[
\Sigma _{13}=\Sigma _3=\Sigma _{-}=0,\quad \Sigma _{+}=-h_1\sqrt{-N_2N_3}%
-1,\quad \Sigma _1=\frac{\sqrt{3}}6\left( N_3-N_2\right) 
\]
\vspace{0.5cm} 
\[
v_1=1,\quad q=2\left( 1+h_1\sqrt{-N_2N_3}\right) ,
\]
\vspace{0.5cm} 
\[
\Omega =\frac{-12h_1\sqrt{-N_2N_3}+12h_1^2N_2N_3-\left( N_2-N_3\right) ^2}%
6,\quad 0<\gamma <2
\]
where $h_1<0$.

\subsection{The exact form of the self-similar equilibrium points}

Apart from the cases with extreme tilt which are not representing by exact
solutions of the EFE, the rest equilibrium points, whenever exist, correspond
to self-similar models i.e. models admitting a proper Homothetic Vector
Field (HVF) $\mathbf{H}$ acting \emph{simply transitively} on space-time: 
\begin{equation}
\mathcal{L}_{\mathbf{H}}g_{ab}=2\psi g_{ab}  \label{symmetry1}
\end{equation}
where $\psi =$const. is the homothetic factor which represents the
(constant) scale transformation of the geometrical and dynamical variables.

For vacuum and non-tilted models, the corresponding self-similar solutions
are all known (see e.g. Tables 9.1. and 9.2., pages 187-188 of \cite
{Wainwright-Ellis}) whereas for tilted models the known self-similar
solutions are of type II \cite{Hewitt-Bridson-Wainwright} and VI$_0$ \cite
{Apostol12,Rosquist-Jantzen1}. Taking into account the results of the
previous section we conclude that only in type VI$_h$ (i.e. $h\neq -1$)
there exists a self-similar model. In order to find the exact form of this
new family of solutions we follow the procedure that was used in \cite
{Apostol9} (a complete study of the self-similar SH models of class B is 
reported elsewhere \cite{Apostol15}).

Adopting the notation of \cite{Ryan-Shepley1}, the KVFs $\{X_\alpha \}$ and
the dual basis $\{\mathbf{\omega }^\alpha \}$ are:

\[
\mathbf{X}_1=\partial _y,\qquad \mathbf{X}_2=\partial _z,\qquad \mathbf{X}%
_3=\partial _x+y\partial _y+h_Rz\partial _z. 
\]
\[
\mathbf{\omega }^1=e^{-x}dy,\qquad \mathbf{\omega }^2=e^{-h_Rx}dz,\qquad 
\mathbf{\omega }^3=dx. 
\]
where the group parameter $h_R$ is formally related with $h$ via ($h_1<0$):\ 
\begin{equation}
h_R=\pm\frac{1-h_1}{1+h_1}.  \label{groupparameters}
\end{equation}
It follows from the non-vanishing structure constants of the isometry group $%
C_{13}^1=h_R^{-1}C_{23}^2=1$ and the Jacobi identities, that the remaining
non-vanishing structure constants $C_{\beta 4}^\alpha $ of the simply
transitive homothety group are:

\[
C_{14}^1=\frac{h_R\left( 2-p_1\right) }{p_2-2},\qquad C_{24}^2=1 
\]
and the HVF is: 
\begin{equation}
\mathbf{H}=\frac{h_R}{p_2-2}t\partial _t+\frac{h_R\left( p_1-2\right) }{2-p_2%
}y\partial _y+z\partial _z.  \label{homothety}
\end{equation}
In addition using the fact that in every perfect fluid model, the fluid
velocity $u^a$ is conformally mapped by the HVF i.e. $\mathcal{L}_{\mathbf{H}%
}u_a=\psi u_a$ in conjunction with the EFE, we write out explicitly the frame
components of the four-velocity and the self-similar metric:\vspace{0.2cm}

\noindent \textbf{Fluid velocity}

\begin{equation}
\Delta _1=v_1t^{p_1-1},\qquad \Delta _2=0,\qquad \Delta _3=v_3t
\label{fluidvelocity}
\end{equation}
\vspace{0.2cm}

\noindent \textbf{Metric}

\begin{equation}
g_{\alpha \beta }=\left( 
\begin{array}{ccc}
c_{11}t^{2(p_1-1)} & 0 & c_{13}t^{p_1} \\ 
0 & c_{22}t^{2(h_R-p_2+2)/h_R} & 0 \\ 
c_{13}t^{p_1} & 0 & c_{33}t^2
\end{array}
\right)  \label{homotheticmetric}
\end{equation}
where $u_a=\Gamma \left( -\delta _a^t+\Delta _\alpha \mathbf{\omega }%
_a^\alpha \right) $ and $\mathbf{g}=-dt^2+g_{\alpha \beta }\mathbf{\omega }%
_a^\alpha \mathbf{\omega }_b^\beta dx^adx^b$.

The various integration constants appearing in (\ref{fluidvelocity}) and (%
\ref{homotheticmetric}) are given by the following expressions:\ 
\[
\gamma =\frac 2{2s+1},\quad p_1=-\frac{2h_R^2\left( s-1\right) +2h_R\left(
2s-1\right) +p_2-2}{h_R^2} 
\]
\vspace{0.5cm} 
\[
\frac {c_{33}}{c_{13}}=\frac{c_{11}(h_R+2)(2s+1)\left[ 2h_R^2s+2h_R\left( 2s-1\right)
+p_2-2\right] +4\tilde{\mu}_0h_R^2v_1\Gamma }{4c_{11}\tilde{\mu}%
_0h_R^2v_1\Gamma } 
\]
\vspace{0.5cm} 
\[
\Gamma^2 =\frac{c_{11}(2s+1)\left[ h_R^2\left( p_2+2s-2\right) +2h_R\left(
2s-1\right) +p_2-2\right] }{2\tilde{\mu}_0h_R^2\left( c_{11}v_3-c_{13}v_1\right) } 
\]
\vspace{0.5cm} 
\[
v_3=-\frac{h_Rv_1^2\left[ 2h_R^2s-2h_R\left( p_2-2s-1\right) +p_2-2\right] }{%
2c_{11}s\left( h_R+2\right) \left[ 2h_R^2s+2h_R\left( 2s-1\right)
+p_2-2\right] } 
\]
\vspace{0.5cm} 
\begin{eqnarray*}
\tilde{\mu}_0 &=&\left( 2s+1\right) \{4h_R^5s\left( 2s-1\right) \left(
2p_2+2s-3\right) - \\
&& \\
&&-2h_R^4\left[ p_2^2\left( 5s-1\right) -p_2\left( 24s^2-3s-1\right)
-2s\left( 20s^2-32s+9\right) \right] + \\
&& \\
&&+h_R^3\left[ p_2^2+4p_2\left( 12s^2-10s+1\right) +4\left(
32s^3-48s^2+20s-1\right) \right] + \\
&& \\
&&+h_R^2\left[ p_2^2\left( 16s-5\right) +12p_2\left( 4s^2-6s+1\right)
+4\left( 16s^3-40s^2+24s-1\right) \right] + \\
&& \\
&&+2h_R\left( p_2-2\right) \left[ p_2\left( 2s+1\right) +2\left(
8s^2-6s-1\right) \right] +4s\left( p_2-2\right) ^2\}\times \\
&& \\
&&\times \left\{ 2h_R^3\left( h_R+2\right) \left( s-1\right) \left[
2h_R\left( p_2-1\right) -p_2+2\right] \right\} ^{-1}
\end{eqnarray*}
\vspace{0.5cm} 
\[
v_1=-\frac{c_{13}\left( h_R+2\right) \left[ 2h_R\left( 2s-1\right) +p_2-2\right]
\left[ 2h_R^2s+2h_R\left( 2s-1\right) +p_2-2\right] }{2h_R^2\left[
h_R^2\left( p_2+2s-2\right) +2h_R\left( 2s-1\right) +p_2-2\right] } 
\]
\vspace{0.5cm} 
\begin{eqnarray*}
c_{11} &=&\left[ 2h_R\left( 2s-1\right) +p_2-2\right] \left[
2h_R^2s+2h_R\left( 2s-1\right) +p_2-2\right] ^2\times \\
&& \\
&&\times \{4h_R^6s\left( 2s-1\right) \left[ p_2\left( s+1\right)
+2s^2-2s-1\right] - \\
&& \\
&&-2h_R^5[p_2^2\left( 3s^2+1\right) -p_2\left( 10s^3+18s^2-13s+5\right) - \\
&& \\
&&-2\left( 24s^4-34s^3+4s^2+5s-2\right) ]-h_R^4[2p_2^3\left( s-1\right) + \\
&& \\
&&+p_2^2\left( 8s^2-24s+15\right) -4p_2\left( 2s^3+18s^2-28s+11\right) - \\
&& \\
&&-4\left( 48s^4-76s^3+24s^2+14s-7\right) ]-h_R^3[3p_2^3\left( s-1\right) +
\\
&& \\
&&+p_2^2\left( 4s^2-38s+23\right) -4p_2\left( 12s^3-28s+13\right) - \\
&& \\
&&-4\left( 32s^4-88s^3+60s^2-5\right) ]+2h_R^2\left( 2-p_2\right) \times \\
&& \\
&&\times [p_2\left( 6s^2-13s+4\right) -2\left( 24s^3-34s^2+13s-2\right) ]+ \\
&& \\
&&+h_R\left( p_2-2\right) ^2[2\left( 12s^2-13s+3\right) -3p_2\left(
s-1\right) ]+ \\
&& \\
&&2\left( p_2-2\right) ^3\left( s-1\right) \}\times \\
&& \\
&&\times \{8h_R^5s\left( s-1\right) [h_R^2\left( p_2+2s-2\right) +2h_R\left(
2s-1\right) +p_2-2]^3\}^{-1}
\end{eqnarray*}
\vspace{0.5cm} 
\begin{eqnarray*}
p_2 &=&2\{h_R^2\left| \left( h_R+2\right) \left[ h_R\left( 2s-1\right)
-s\right] \right| \times \\
&& \\
&&\times [\left( 2s-1\right) (h_R^2\left( s+1\right) ^2\left( 2s-1\right)
+2h_R\left( 4s^3-4s^2+5s-1\right) + \\
&& \\
&&+8s^3-28s^2+10s-1)]^{1/2}+h_R^5\left( s+1\right) \left( 2s-1\right) ^2- \\
&& \\
&&-h_R^4\left( 2s^3-21s^2+17s-4\right) -h_R^3\left( 8s^3+9s-4\right) + \\
&& \\
&&+h_R^2\left( 20s^3-16s^2+4s-1\right) -2h_R\left( 4s^3+6s^2-8s+1\right)
+4s^2\}\times \\
&& \\
&&\times \{2h_R^4\left( 8s^2-5s+1\right) -h_R^3\left( 16s^2+6s-5\right) + \\
&& \\
&&+h_R^2\left( 20s^2-14s+1\right) -2h_R\left( 8s^2-8s+1\right) +4s^2\}^{-1}
\end{eqnarray*}
\newline
\newline
From the expressions of the constants $p_2$, $\Gamma $ and the positivity of
the energy density it can be verified that the above family of self-similar
tilted perfect fluid solutions of type VI$_h$ is defined for 
\begin{equation}
\frac 16<s<\frac{h_R}{2h_R-1}\Leftrightarrow \frac{2\left( 3-h_1\right) }{%
5-3h_1}<\gamma <\frac 32  \label{bound1}
\end{equation}
when $-1<h_R<-\frac 12$ and

\begin{equation}
\frac{h_R}{2h_R-1}<s<\frac 12\Leftrightarrow 1<\gamma <\frac{2\left(
3-h_1\right) }{5-3h_1}  \label{bound2}
\end{equation}
when $-\frac 12<h_R<0$. We note that the inequalities involving the bounds
of the group parameter $h_R$ are reduced to $-\frac 13<h_1<0$ and $%
-1<h_1<-\frac 13$ respectively as expected from the analysis in the previous
section.

As a concrete and illustrative example of the general results given above,
we present the self-similar metric and the associated energy density for the
case of a radiation fluid i.e. $\gamma =\frac 43\Leftrightarrow s=\frac 14$: 
\[
p_1=-\frac{\left| \left( h_R+2\right) \left( 2h_R+1\right) \right| \sqrt{%
25h_R^2-4h_R+4}-12h_R^4-58h_R^3+7h_R^2-32h_R-4}{4\left(
2h_R^4+10h_R^3-5h_R^2+4h_R+1\right) } 
\]
\[
p_2=\frac{\left| \left( h_R+2\right) \left( 2h_R+1\right) \right| \sqrt{%
25h_R^2-4h_R+4}+10h_R^5+33h_R^4+52h_R^3-22h_R^2+36h_R+8}{4h_R^{-2}\left(
2h_R^4+10h_R^3-5h_R^2+4h_R+1\right) } 
\]
\begin{eqnarray*}
\tilde{\mu} &=&\{-3h_R\left| h_R+2\right| \left(
6h_R^4-5h_R^3+6h_R^2+6h_R+2\right) \sqrt{25h_R^2-4h_R+4}+ \\
&& \\
&&+30h_R^5-29h_R^4+16h_R^3-30h_R^2-10h_R+2\}\times \\
&& \\
&&\times \{4t^2\left( 2h_R^4+10h_R^3-5h_R^2+4h_R+1\right) [-h_R\left(
2h_R-1\right) \sqrt{25h_R^2-4h_R+4}+ \\
&& \\
&&+10h_R^3-5h_R^2+6h_R+2]\}^{-1}.
\end{eqnarray*}
It turns out that the energy density is increasing as the group parameter
varies in the interval $h_R\in \left( -1,-\frac 12\right) $ or in terms of
the ``original'' parameter $h\in \left( -\frac 19,0\right) $.

\section{Concluding Remarks}

\setcounter{equation}{0}

A long term goal of the qualitative study of SH tilted perfect fluid models
rely on the determination of the equilibrium points of their dynamical state
space in order to be able to make solid conjectures regarding the dynamics,
at the asymptotic regimes, of the corresponding models. In the present
paper, by exploiting the orthonormal frame formalism and the general form of
evolution equations given in \cite{Hewitt-Bridson-Wainwright}, we have found
the complete set of equations describing the evolutionary behaviour of the
important class of Bianchi type VI$_h$ tilted perfect fluid models.

As a result we have identified the dynamical state space to be a bounded
region $\mathcal{D}\subset \mathbf{R}^{7}$ subjected to the constraints (%
\ref{constraints1})-(\ref{constraints3}) and we have given all the
equilibrium points (both vacuum and non-vacuum). This study led us to a new
family of exact solutions of EFE which are represented by the self-similar
metrics (\ref{homotheticmetric}). This family of \emph{rotational} tilted
perfect fluid models, belongs to the subclass $n_\alpha ^\alpha =0$ and has
some interesting stability properties.

In particular we have seen that this model is defined only for $-1<h<0$ and
has a five dimensional stable manifold in the ranges $h_1<-\frac
13\Rightarrow 1<\gamma <\frac{2\left( 3-h_1\right) }{5-3h_1}$ and $-\frac
13<h_1<0\Rightarrow \frac{2\left( 3-h_1\right) }{5-3h_1}<\gamma <\frac 32$.
In conjunction with the fact that for $-1<h<0$ the Collins non-tilted
solution $\mathcal{C}^{\pm }(VI_h)$\textit{\ }is stable whenever $\frac
23<\gamma \leq \frac{2\left( 3-h_1\right) }{5-3h_1}$ \cite{Barrow-Hervik} we
conclude that, at the value\footnote{%
We note that, in contrast with the case of VI$_0$ models in which a similar
situation occurs concerning the value $\gamma =\frac 65$, there is no
acceptable tilted solution for the value $\gamma =\frac{2\left( 3-h_1\right) 
}{5-3h_1}$.} $\gamma =\frac{2\left( 3-h_1\right) }{5-3h_1}$ and when $%
1+3h_1>0$, there is an exchange of stability between the Collins model and
the present solution. Moreover for $1+3h_1<0$ both models are stable
whenever $1<\gamma <\frac{2\left( 3-h_1\right) }{5-3h_1}$ and they are also
destabilised at $\gamma =\frac{2\left( 3-h_1\right) }{5-3h_1}$. However, due
to the complexity of the situation, there are no straightforward conclusions
regarding the asymptotic behaviour of the models in the full state space.
Perhaps a more sophisticated choice of the state variables is needed in
order to efficiently answer many open questions. Nevertheless the case $h=-1$
appears to be more tractable, permitting a complete analysis of its
dynamical properties. We also pointed out that the above procedure can be
used to determine the equilibrium points for the rest of the Bianchi models
and for the case of the exceptional models $h=-\frac 19$ as well. These
matters will be the subject of a future work.

A quick glance of the equilibrium points in section 3 shows the lack of
physically acceptable models for $h=-1$ which proves (this fact can be
confirmed using the geometric results of \cite{Apostol15}) the non-existence
of tilted self-similar type III models. Furthermore in all (non extreme
tilted) cases, the tilted fluid velocity is not parallel to $A_\alpha $.
According to Theorem 3.2. of \cite{King-Ellis} this implies that
self-similar type VI$_h$ models have necessary non-zero vorticity. In
conclusion we have the following proposition:\newline
\newline
\textbf{Proposition 1.} \emph{There are no type III and irrotational type VI$%
_h$ tilted perfect fluid models which admit a proper HVF. }\newline
\newline
We finally note that a stability analysis of the irrotational tilted perfect
fluid models of type VI$_h$ shows that the equilibrium points (both
non-tilted and tilted) found in section 3 are future unstable which, by
means of Proposition 1, implies that \emph{irrotational type VI$_h$ models
are not asymptotically self-similar but rather they are asymptotic to an
extreme tilted model}.

\vskip 0.5cm

\centerline{\bf\large Acknowledgments} \vskip .3cm \noindent The author
gratefully acknowledges the hospitality of the Universitat de les Illes
Balears, Spain (UIB) where part of this work was carried out and Jaume Carot
for useful discussions. He also acknowledges the financial support of
Ministry of National Education and Religious Affairs through the research
program ``Pythagoras'', grant No 70-03-7310.

\end{document}